\documentclass[11pt,a4paper]{article}
\usepackage{jheppub_kim}
\usepackage{rotating} 
\usepackage{graphicx,epsfig}
\usepackage{amsmath}
\usepackage {amssymb}
\usepackage{subfigure}

 \def\e{\mathrm{e}}

\usepackage{relsize}

\usepackage{graphicx}
\usepackage{epsfig}
\usepackage{bm}
\usepackage{amssymb}
\usepackage{float}
\usepackage{amsmath}
\usepackage{dcolumn}
\usepackage{cancel}

\providecommand{\U}[1]{\protect\rule{.1in}{.1in}}

\newcommand{\newc}{\newcommand}

\newc{\be}{\begin{equation}}
\newc{\ee}{\end{equation}}

\newc{\ba}{\begin{eqnarray}}
\newc{\ea}{\end{eqnarray}}
\newc{\bea}{\begin{eqnarray*}}
\newc{\eea}{\end{eqnarray*}}
\newc{\D}{\partial}
\newc{\ie}{{\it i.e.} }
\newc{\eg}{{\it e.g.} }
\newc{\etc}{{\it etc.} }
\newc{\etal}{{\it et al.}}
\newc{\lcdm}{$\Lambda$CDM }

\newc{\ra}{\Rightarrow}

\title{Holographic dark energy through Loop Quantum Gravity inspired entropy}

\author{Andreas Lymperis}
 
\affiliation{Department of Physics, University of Patras, 26500 Patras, 
Greece}

\emailAdd{alymperis@upatras.gr}

\abstract{ 
We construct a new cosmological holographic dark energy scenario based on Loop Quantum Gravity inspired entropy, instead of the standard  Bekenstein-Hawking one. The former is an extended black-hole entropy that arises from non-extensive statistics and quantum geometry and is quantified by a new dimensionless parameter $q$, which possesses standard holographic dark energy as a  particular sub-case. In the future event horizon as the Infrared cutoff, we provide a simple differential equation for the dark energy density parameter, as well as an analytical expressions for the corresponding equation-of-state and deceleration parameters. We show that the scenario at hand can describe successfully the usual thermal history of the Universe, with the sequence of matter and dark-energy epochs, while the transition to acceleration takes place at $z\approx 0.6$. Additionally, according to the value of the new entropic parameter $q$, the dark energy equation-of-state parameter can have 
a rich behavior, and it can be quintessence-like, phantom-like, or experience the phantom-divide crossing.}

\begin{document}
\maketitle

\section{Introduction}
Although General Relativity (GR) and Standard Model of Particle Physics have an undoubtedly success in describing the gravitational and particle interactions respectively, they cannot easily describe the early-time 
and late-time accelerated epochs of the Universe, according to the concordance paradigm of cosmology, \ie $\Lambda$CDM scenario, and thus the necessity of extra degree(s) of freedom it seems to be required. This leads one to either attribute these extra degrees of freedom to new, exotic forms of matter, 
such as in the dark energy concept  \cite{Copeland:2006wr,Cai:2009zp,Bamba:2012cp}, or alternatively can attribute these extra degrees of freedom, to gravitational modifications that possesses GR as a particular limit, namely to have gravitational origin \cite{Nojiri:2006ri,Nojiri:2010wj,Capozziello:2011et,Cai:2015emx,Nojiri:2017ncd,Asimakis:2022jel}.

Beyond the two main aforementioned directions, an interesting alternative scenario for the explanation of dark energy origin and nature, is the holographic dark energy, which is originated from application of the fundamental holographic principle \cite{tHooft:1993dmi,Susskind:1994vu,Bousso:2002ju} at a cosmological framework \cite{Fischler:1998st,Bak:1999hd,Horava:2000tb}. The crucial ingredient for the quantum field theory applicability at large distances, is the connection between the Ultraviolet (UV) cutoff of a quantum field theory (which is related to the vacuum energy), and the largest length of this theory \cite{Cohen:1998zx}. The resulting vacuum energy will be a form of dark energy of holographic origin, namely holographic dark energy (HDE) \cite{Li:2004rb,Wang:2016och}. Both its simple versions \cite{Li:2004rb,Wang:2016och,Horvat:2004vn,Huang:2004ai,Pavon:2005yx,Wang:2005jx,Nojiri:2005pu,Kim:2005at,Wang:2005ph, Setare:2006wh,Setare:2008pc,Setare:2008hm}, as well as its extended versions \cite{Saridakis:2007cy,Setare:2007we,Cai:2007us,Setare:2008bb,Saridakis:2007ns,Saridakis:2007wx,Jamil:2009sq,Gong:2009dc,Suwa:2009gm,BouhmadiLopez:2011xi,Chimento:2011pk,Malekjani:2012bw,Chimento:2013qja,Khurshudyan:2014axa,
Landim:2015hqa,Pasqua:2015bfz,Jawad:2016tne,Pourhassan:2017cba,Nojiri:2017opc,Saridakis:2017rdo,
Saridakis:2018unr,Aditya:2019bbk,Nojiri:2019kkp,Geng:2019shx,Waheed:2020cxw} have been proven to lead to interesting cosmological phenomenology, and additionally to be in agreement with observational data \cite{Zhang:2005hs,Li:2009bn,Feng:2007wn,Zhang:2009un,Lu:2009iv,Micheletti:2009jy,DAgostino:2019wko,Sadri:2019qxt,Molavi:2019mlh}.

Moreover, the construction of holographic dark energy can be achieved through the connection of the entropy of a system  with geometrical quantities such as its radius. As it is widely known, the black-hole and cosmological application of the standard Boltzmann-Gibbs entropy has as a result the standard Bekenstein-Hawking entropy. Since we are interested to the cosmological application of holography, it is necessary to extend the standard Boltzmann-Gibbs theory to a nonextensive one. This necessity of a generalized entropy is resulting from the fact that the Universe horizon, \ie the largest distance is proportional to its area, similarly to a black hole. In this context, Gibbs had pointed out that in systems in which the partition function diverges the Boltzmann-Gibbs theory cannot be applied, and the gravitational systems lie within this class. Thus, in order to apply holography and entropy relations to the whole Universe, which is a gravitational and thus nonextensive system, one should use a generalized definition of the universe horizon entropy. In this context, several generalized entropies have been proposed such as Tsallis entropy \cite{Tsallis:1987eu,Lyra:1998wz}, Barrow entropy \cite{Barrow:2020tzx}, Kaniadakis entropy \cite{Kaniadakis:2002zz,Kaniadakis:2005zk}, Nojiri-Odintsov-Paul entropy \cite{Nojiri:2022dkr}, and their implications in black-hole thermodynamics \cite{Lenzi:1998br,Jawad:2022lww,Abreu:2020wbz,Abreu:2020rrh,Abreu:2020dyu}, and in cosmology \cite{Lymperis:2018iuz,Lymperis:2021qty,Basilakos:2023kvk,Nojiri:2019skr,Lymperis:2021emd,Cai:2008ys,Sheykhi:2018dpn,Sheykhi:2023aqa,Sheykhi:2019bsh,Saridakis:2020lrg,Nojiri:2022aof,Odintsov:2022qnn,Odintsov:2023qfj,Luciano:2022knb,Luciano:2023bai,Luciano:2023zrx,Luciano:2023wtx,Luciano:2023roh,Luciano:2022hhy,Luciano:2022eio,Ghaffari:2022skp,Luciano:2022pzg,Moradpour:2020dfm,Hernandez-Almada:2021rjs,Sarkar:2021izd,Nojiri:2021jxf,Asghari:2021bqa,Paul:2022doh,Abreu:2017hiy,Drepanou:2021jiv,Abreu:2018phv,Abreu:2017fhw,Jawad:2021xsr,Abreu:2018mti,Leon:2021wyx,Barrow:2020kug,Moradpour:2020kss,Koussour:2022sdy,Ghaffari:2021xja,Barboza:2014yfe,Abreu:2012msk,Lima:2001vq,AbdollahiZadeh:2018ubg,Huang:2019hex,Zhang:2019zxv,Anagnostopoulos:2020ctz,Saridakis:2020zol,Saridakis:2020cqq,Hernandez-Almada:2021aiw,Huang:2021zgj,Mamon:2020spa,Mohammadi:2021wde,Jawad:2021xsr,Nojiri:2021czz,Aly:2019otq,Kohyama:2006hu,Bhattacharjee:2020rqk,daSilva:2020bdc,Saha:2020vxn,Abreu:2017fhw,Mamon:2020wnh,Jawad:2019ouc,Ghaffari:2019qcv,Abreu:2017hiy,Telali:2021jju}. However, an interesting generalization definition of the Universe horizon entropy has been inspired from the Loop Quantum Gravity where the defined entropic parameter quantifies how the probability of frequent events is enhanced relatively to infrequent ones \cite{Majhi:2017zao,Veraguth:2017uwp,Liu:2021dvj}. In such an extended non-extensive theory, the distribution functions are a one-parameter continuous deformation of the usual Maxwell-Boltzmann ones, and hence standard statistical theory is recovered in a particular limit.

In the present work we are interested to construct a new cosmological scenario resulting as a consistent formulation of Loop Quantum Gravity inspired entropy and investigate its cosmological implications. The resulting scenario is a consistent generalization of standard holographic dark energy, possessing it as a particular limit \ie when Loop Quantum entropy becomes the standard Bekenstein-Hawking one. The plan of the manuscript is the following: In Section \ref{model} we 
formulate Loop Quantum Gravity holographic dark energy in a consistent way, extracting analytical relations for the corresponding cosmological equations. In Section \ref{cosmoevo} we study the cosmological behavior of the scenario, focusing on the evolution of the dark energy density and equation-of-state parameters. Finally, in Section \ref{concl} we summarize our results.

\section{Loop Quantum Gravity holographic dark energy} \label{model}
In this section we desire to formulate LQG holographic dark energy in a consistent way, in order to investigate its cosmological implications. First, let us briefly review some basic elements of non-extensive statistical mechanics and the structures of the quantum geometry of black holes in Loop Quantum gravity.

Considering the deviations on the probability of a given microstate of the potential quantum mechanical system (\ie the measure of the departure from additivity), Tsallis was first introduced the $q$-entropy or as it is widely-known Tsallis entropy, which has the form \cite{Tsallis:1987eu,Lyra:1998wz}
\be \label{tsallisentr}
S_{q}=k_{B}\frac{\left (1-\sum^{W}_{i=1}{p^{q}_{i}}\right )}{q-1},
\ee
where $q$ is the entropic index. In the case where $q > 1$, frequent events with probability close to 1 are relatively enhanced, while in case where $q < 1$ rare events with a probability close to zero are relatively enhanced. For equal probability, \ie $p_{i}=1/\Omega$ for all $i$, equation (\ref{tsallisentr}) reduces to 
\be \label{eqprob}
S_{q}=k_{B}\ln_{q}\Omega ,
\ee
where $\ln_{q}(x)$ is the $q$-logarithm. Furthermore, considering $N$ particles with spin (${j_{1},j_{2},\dots,j_{N}}$), the $\Omega$ parameter in equation (\ref{eqprob}) takes the form $\Omega (j_{1},j_{2},\dots,j_{N})=\Pi^{N}_{u=1} (2j_{u}+1)$ and with $j_{1},j_{2},\dots,j_{N}=s$ equation (\ref{eqprob}) can be written as
\be \label{sqform}
S_{q}=k_{B}\frac{\left (1+2s\right)^{(1-q)^{N}}-1}{1-q}.
\ee

Moreover, in the LQG framework the structure of the quantum geometry of black holes horizon  cross-section can be described by a topological two-sphere with defects (punctures), in which their spin quantum number can be defined by the edge of the spin network, which represents the bulk quantum geometry \cite{Majhi:2017zao,Mejrhit:2019oyi}. Furthermore, in order to apply statistical mechanics into black hole thermodynamics framework, the number of punctures $N$ must be very large, \ie $N\rightarrow \infty$, and additionally, the minimal quantum of area can be determinced when spin is $1/2$. Thus, the number of possible states can be derived by $j_{i}=1/2$. Therefore, applying all these into equation (\ref{sqform}), the q-entropy of a black hole can be written as \cite{Majhi:2017zao,Mejrhit:2019oyi}
\be \label{sqformbh} 
S_{q}=k_{B}\frac{2^{(1-q)^{N}}-1}{1-q},
\ee
with the area of the black hole to be given by
\be \label{bharea}
A=4\pi \gamma L^{2}_{p}N\sqrt{3}.
\ee
Using equation (\ref{sqformbh}) and (\ref{bharea}) the black hole entropy in LQG cosmology can be written as
\be \label{lqgentr}
S_{LQG}=\frac{k_{B}}{1-q}\left (\e^{(1-q)\Lambda (\gamma_{0})S_{BH}}-1\right ),
\ee
where the parameter $\Lambda (\gamma_{0})$ is defined as
\be 
\Lambda (\gamma_{0})=\frac{\ln{2}}{\sqrt{3}\pi \gamma_{0}},
\ee
$S_{BH}=k_{B}\frac{A}{4G}$ is the usual Bekenstein-Hawking entropy, and $\gamma_{0}$ is the Barbero-Immirzi parameter. Depending on the gauge group used in LQG the parameter $\gamma_{0}$ can be take one of the two possible values $\frac{\ln{2}}{\pi \sqrt{3}}$ or $\frac{\ln{3}}{2\pi \sqrt{2}}$, but in general is a free parameter in scale-invariant gravity \cite{Veraguth:2017uwp,Wang:2019ryx}. In the case where $\gamma_{0}=\frac{\ln{2}}{\pi \sqrt{3}}$, the parameter $\Lambda(\gamma_{0})=1$ and in the limit $q\rightarrow 1$, equation (\ref{lqgentr}) is reduced  to the standard  Bekenstein-Hawking entropy one. From now on we will work in natural units in which $k_{B} = c = \hbar = 1$.

Since when $q\rightarrow 1$ (LQG entropy is close to the usual Bekenstein-Hawking one), the term $1-q \rightarrow 0$, and hence it is justified to expand LQG entropy, namely equation (\ref{lqgentr}), for small values of the term $1-q$ obtaining
\be \label{lqgentr1}
S_{LQG} = \Lambda(\gamma_{0})S_{BH}+\frac{(1-q)\Lambda^{2}(\gamma_{0})S^{2}_{BH}}{2}+ {\cal{O}}[(1-q)^{3}].   
\ee

With equation (\ref{lqgentr1}) at hand we can now proceed to the formulation of LQG holographic dark energy. The standard holographic dark energy is given by the inequality $\rho_{DE}L^4\leq S$, with $L$ to be the the horizon length (the largest distance of 
the theory, \ie the Infrared cutoff), and $S$ the entropy relation applied in a 
black hole of radius $L$ \cite{Li:2004rb,Wang:2016och} (see also \cite{10.1088/1572-9494/acf27c} and the references therein). Hence, inserting (\ref{lqgentr1}) into the aforementioned inequality and imposing $S\propto A\propto L^{2}$, LQG holographic dark energy can be written as
\be \label{lqghde}
\rho_{DE}= 3c^{2} M_p^{2} \Lambda(\gamma_{0}) L^{-2}+ 3\tilde{c}^{2}(1-q) M_p^{4} \Lambda^{2}(\gamma_{0}),
\ee
where $M_p$ the Planck mass and $c$ and $\tilde{c}$ are constants. In the case where $q\rightarrow 1$, equation (\ref{lqghde}) reduces to the standard holographic dark energy  $\rho_{DE}=3c^2 M_p^2 L^{-2}$. In what follows we absorb the constant  $\tilde{c}$ inside the term $(1-q)$, by setting $3\tilde{c}^{2}(1-q)\equiv\widetilde{(1-q)}$ and we drop the tildes for simplicity.

We proceed now by considering a flat homogeneous and isotropic Friedmann-Robertson-Walker (FRW) Universe which is defined by  the metric  
\be \label{FRWmetric}
ds^{2}=-dt^{2}+a^{2}(t)\delta_{ij}dx^{i}dx^{j}\,,
\ee
where $a(t)$ is the scale factor. As it mentioned in the previous paragraph, $L$ is the Infrared cutoff, which it is needed to be defined in any holographic dark energy scenario, in order one to study in detail the scenario at hand. In this context,                                                                                                                                                                                                                                                            the choice of the Hubble horizon $H^{-1}$ is not suitable, since this choice leads to inconsistencies \cite{Hsu:2004ri}, such as no acceleration. Therefore, as it was shown, one must use the future event horizon, which is given by the relation \cite{Li:2004rb} 
\be \label{futrhoriz}
R_h\equiv a\int_t^\infty \frac{dt}{a}= a\int_a^\infty \frac{da}{Ha^2}.
\ee

Having defined the Infrared cutoff, the energy density of LQG holographic dark energy (\ref{lqghde}) can be written as
\be \label{lqghde1}
\rho_{DE}= 3c^{2} M_p^{2} \Lambda(\gamma_{0}) R_{h}^{-2}+(1-q) M_p^{4} \Lambda^{2}(\gamma_{0}).
\ee
In a Universe filled with dark energy and matter perfect fluids, the Friedmann equations are
\begin{eqnarray}
\label{Fr1b}
H^2& =& \ \frac{1}{3M_p^2}\left (\rho_m + \rho_{DE}\right )    \\
\label{Fr2b}
\dot{H}& =& -\frac{1}{2 M_p^2}\left (\rho_m +p_m+\rho_{DE}+p_{DE}\right),
\end{eqnarray}
with $p_{DE}$ the pressure of LQG holographic dark energy, and $\rho_m$ 
and $p_m$ the energy density and pressure of the matter sector respectively.
The set of equations close by considering the matter conservation equation  
\be \label{rhoconserv}
\dot{\rho}_m+3H(\rho_m+p_m)=0.
\ee
Finally, it proves convenient to introduce the dark energy and matter density parameters 
 \begin{eqnarray}
 && \Omega_m\equiv\frac{1}{3M_p^2H^2}\rho_m
 \label{Omm}\\
 &&\Omega_{DE}\equiv\frac{1}{3M_p^2H^2}\rho_{DE}.
  \label{ode}
 \end{eqnarray}
 
Combining equations (\ref{futrhoriz}),(\ref{lqghde1}) and (\ref{ode}) we obtain
\be \label{integrrelation}
\int_x^\infty \frac{dx}{Ha}=\frac{1}{a}\left( \frac{3c^{2}\Lambda_{\gamma_{0}}}{3H^2 \Omega_{DE}-(1-q)M_p^{2}\Lambda^{2}_{\gamma_{0}}}
\right)^{\frac{1}{2}},
\ee
with $x\equiv \ln a$, and from (\ref{lqghde1}) we have kept only the solutions that give positive $R_h$ and 
additionaly it reduces to the usual result $\int_x^\infty \frac{dx}{Ha}=\frac{c}{aH\sqrt{\Omega_{DE}}}$ when $q\rightarrow 1$ and for $\Lambda({\gamma_{0})=1}$. 

Furthermore, we focus on the case where the Universe is filled with dust matter, in which case the pressure of matter sector is set to zero ($p_{m}=0$) and consequently this leads to $w_{m}=0$. Hence, in this case the conservation equation (\ref{rhoconserv}) leads to $\rho_m=\rho_{m0}/a^3$, where $\rho_{m0}$ is the matter energy density at the current scale factor $a_0=1$ (in what follows the subscript ``0'' will denote the present value of a quantity). Therefore, substituting  into equation (\ref{Omm}) leads to $\Omega_m=\Omega_{m0} H_0^2/(a^3 H^2)$. Additionally, from the first Friedmann equation (\ref{Fr1b}) using the relations (\ref{Omm}) and (\ref{ode}) we have $\Omega_m+\Omega_{DE}=1$ where we can obtain the useful relation
\be \label{Hrel2}
\frac{1}{Ha}=\frac{\sqrt{a(1-\Omega_{DE})}}{H_0\sqrt{\Omega_{m0}}}.
\ee

Substituting equation (\ref{Hrel2}) into equation (\ref{integrrelation}) we obtain
\be \label{integrrelation2}
\int^{\infty}_{x}\frac{dx}{H_{0}\sqrt{\Omega_{m0}}}\sqrt{a(1-\Omega_{DE})}=\frac{1}{a}\left( \frac{3c^{2}\Lambda(\gamma_{0})}{3H^2 \Omega_{DE}-(1-q)M_p^{2}\Lambda^{2}_{\gamma_{0}}}
\right)^{\frac{1}{2}},
\ee
where $x=\ln a$ plays the role of the independent variable, and therefore for a quantity $f$ we always acquire $\dot{f}=f' H$, 
with primes denoting derivatives with respect to $x$. Differentiating (\ref{integrrelation2}) in terms of $x$ we obtain
\be \label{Odediffeq}
\Omega_{DE}'=\Omega_{DE}(1-\Omega_{DE})\left 
\{1+\frac{2Q(1-\Omega_{DE})}{3\Omega_{DE}}+\frac{2\sqrt{3}}{9}\frac{\left[3\Omega_{DE}-Q(1-\Omega_{DE})\right]^{\frac{3}{2}}}{c\Lambda^{\frac{1}{2}}\Omega_{DE}} \right\},
\ee
where
\be
Q=\frac{\e^{3x}(1-q)M^{2}_{p}\Lambda^{2}}{H^{2}_{0}\Omega_{m0}}.
\ee
Equation (\ref{Odediffeq}) determines the evolution of LQG holographic dark energy as a function of $x=\ln a$.
We observe that in the limit when $q\rightarrow 1$ and $\Lambda(\gamma_{0})=1$ (\ie when $\gamma_{0}=\frac{\ln2}{\pi \sqrt{3}}$), equation (\ref{Odediffeq}) recovers the corresponding differential equation of the usual holographic dark energy 
\cite{Li:2004rb}, \ie
$\Omega_{DE}'= 
\Omega_{DE}(1-\Omega_{DE})\left(1+2\frac{\sqrt{\Omega_{DE}}}{c}\right)$, which accepts an analytic solution in an implicit form  \cite{Li:2004rb}. Nevertheless, in the general case where $q\neq 1$, equation (\ref{Odediffeq}) exhibits an explicit $x$-dependence and one cannot obtain an analytical solution. Thus, in the following we will elaborate it numerically.

The next important observable of LQG holographic dark energy scenario to determine is the equation-of-state parameter $w_{DE}\equiv p_{DE}/\rho_{DE}$. Since the matter sector is conserved, using the Friedmann equations (\ref{Fr1b}),(\ref{Fr2b}) into the conservation equation (\ref{rhoconserv}), we deduce that the dark energy sector is conserved too, namely
\be \label{rhodeconserv}
\dot{\rho}_{DE}+3H\rho_{DE}(1+w_{DE})=0.
\ee
In order to extract the relation of $w_{DE}$ in LQG holographic dark energy scenario, we proceed by differentiating the basic equation (\ref{lqghde1}) which gives $\dot{\rho}_{DE}=-6c^{2}M^{2}_{p}\Lambda(\gamma_{0})R^{-3}_{h}\dot{R}_{h}$, where $\dot{R}_h=H  R_h-1$ from (\ref{futrhoriz}). Furthermore, $R_h$ can be further eliminated in terms of $\rho_{DE}$, according to (\ref{lqghde1}), throuth the relation
\be \label{rh}
 R_h=\left 
(\frac{3c^{2}M^{2}_{p}\Lambda(\gamma_{0})}{\rho_{DE}-(1-q)M^{4}_{p}\Lambda^{2}(\gamma_{0})}
\right )^{1/2}.
\ee
Inserting all the above into (\ref{rhodeconserv}) we obtain
\be \label{rhodeconserv2}
-6c^{2}M^{2}_{p}\Lambda(\gamma_{0})R^{-3}_{h}\left [ H\left 
(\frac{3c^{2}M^{2}_{p}\Lambda(\gamma_{0})}{\rho_{DE}-(1-q)M^{4}_{p}\Lambda^{2}(\gamma_{0})}
\right )^{1/2}-1\right ]+3H\rho_{DE}(1+w_{DE})=0.
\ee
Finally, substituting $H$ from (\ref{Hrel2}), and using the dark energy density parameter definition 
(\ref{ode}) we result to
\be \label{wDE}
w_{DE}=-1+\frac{2\left[3\Omega_{DE}-Q(1-\Omega_{DE})\right]\left[3c\sqrt{\Lambda}-\sqrt{3}\sqrt{3\Omega_{DE}-Q(1-\Omega_{DE})}\right]}{27c\sqrt{\Lambda}\Omega_{DE}}.
\ee
Relation (\ref{wDE}) provides the dark energy equation-of-state parameter, as a function of $\ln a$, in LQG holographic dark energy scenario. Since the solution of $\Omega_{DE}$ is given from (\ref{Odediffeq}), the solution of $w_{DE}$ can be given from (\ref{wDE}). As expected when $q\rightarrow 1$ and $\Lambda(\gamma_{0})=1$ the above expression, provides the standard 
holographic dark energy result, namely $w_{DE}=-\frac{1}{3}-\frac{2}{3}\frac{\sqrt{\Omega_{DE}}}{c}$ \cite{Wang:2016och}.

We close this section by introducing the deceleration parameter $q$ which can be written as
\be \label{qdeccel}
q\equiv-1-\frac{\dot{H}}{H^2}=\frac{1}{2}+\frac{3}{2}\left(w_m\Omega_m+w_{DE}\Omega_{DE}\right),
\ee
and in the case of dust matter is straightforwardly known as long as $\Omega_{DE}$ and consequently $w_{DE}$ are known.
 
\section{Cosmological evolution} \label{cosmoevo}
\begin{figure}[!]
\centering
\includegraphics[width=8.cm]{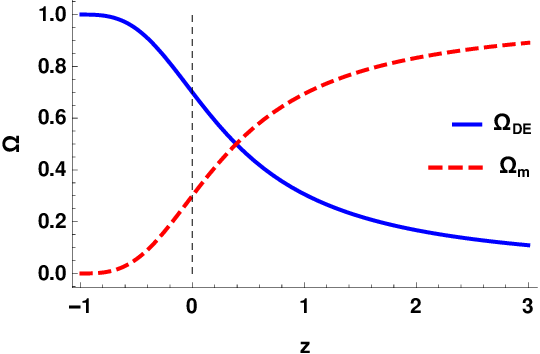}    \\                           
\includegraphics[width=8.cm]{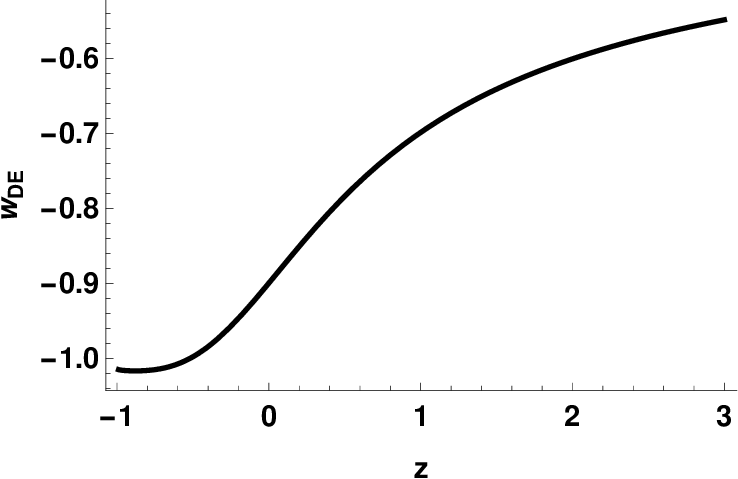} \\
\includegraphics[width=8.cm]{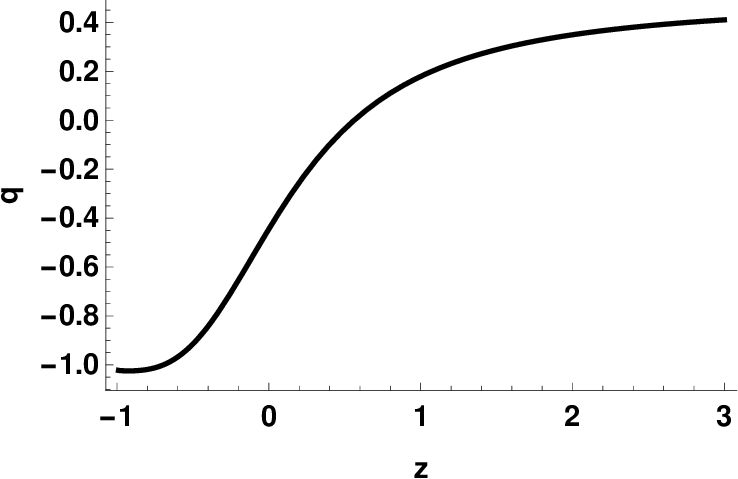}
\caption{\it{ {\bf{Upper graph}}: The LQG holographic dark energy  density   parameter $\Omega_{DE}$ (blue-solid) and the matter density parameter $\Omega_{m}$ (red-dashed), as a function of the redshift $z$, for
$q=1.1$ and $c=1=\Lambda(\gamma_{0})$, in units where $M^{2}_{p} =1$. {\bf{ Middle graph}}: The corresponding dark energy equation-of-state parameter $w_{DE}$. {\bf{Lower graph}}:  The corresponding deceleration parameter $q$. In all graphs we have set $\Omega_{DE}(x=-\ln(1+z)=0)\equiv\Omega_{DE0}\approx0.7$ in agreement with observations.
}}
\label{HDEOmegas}
\end{figure}
\begin{figure}[!h]
\centering
\includegraphics[width=8cm]{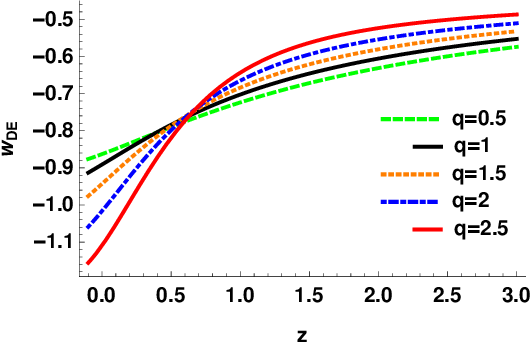}                             
  \caption{\it{ The redshift-evolution of the equation-of-state parameter 
$w_{DE}$ of LQG
holographic dark energy, for fixed $c=1$ and $\Lambda(\gamma_{0})=1$ and various values of $q$ in units where $M^{2}_{p} =1$. We have imposed $\Omega_{DE}(x = -\ln{(1 + z)} = 0) \equiv \Omega_{DE0}\approx 0.7$ at present in agreement with observations.
}}
\label{hdemultiwdeq}
\end{figure} 

In this section we proceed to a detailed study of the cosmological behavior of the resulting holographic scenario of the previous section, namely the scenario of Loop Quantum Gravity holographic dark energy. Elaboration of the differential equation (\ref{Odediffeq})  numerically (since it cannot be solved analytically except for $q\rightarrow 1$) we acquire the solution for $\Omega_{DE}(x)$. Since we are interested to investigate its behavior, it is more convenient to use the redshift parameter $z$ instead of $x$ through the relation $x\equiv\ln a=-\ln(1+z)$. Additionally, imposing $\Omega_{DE}(x=-\ln(1+z)=0)\equiv\Omega_{DE0}\approx0.7$ and therefore $\Omega_m(x=-\ln(1+z)=0)\equiv\Omega_{m0}\approx0.3$ in agreement with observations \cite{Planck:2018vyg}, in the upper graph of Fig. \ref{HDEOmegas} we present the evolution of the dark energy and matter density parameters in terms of the redshift parameter $z$. As we can see the scenario at hand can provide a successfully description of the thermal history of the Universe, with the sequence of matter and dark energy epochs and in the far future, \ie at $z\rightarrow-1$, the Universe results asymptotically to a complete dark-energy dominated phase. Moreover, in the middle graph of  Fig. \ref{HDEOmegas} we depict the corresponding behavior of the dark-energy equation-of-state parameter $w_{DE}$ which can be determined from (\ref{wDE}). From this figure we observe that the value of $w_{DE}$ at present is around $-1$ in agreement with observations and in which case $w_{DE}$  in the future enters slightly inside the phantom regime (which behavior is allowed according to equation (\ref{wDE})), showing the enhanced capabilities of the scenario at hand. Lastly, in the lower-graph of Fig. \ref{HDEOmegas} we depict the deceleration parameter $q$, where we can see that the transition from deceleration to acceleration happens at $z\approx 0.6$ in agreement with observational data.
\begin{figure}[!h]
\centering
\includegraphics[width=8cm]{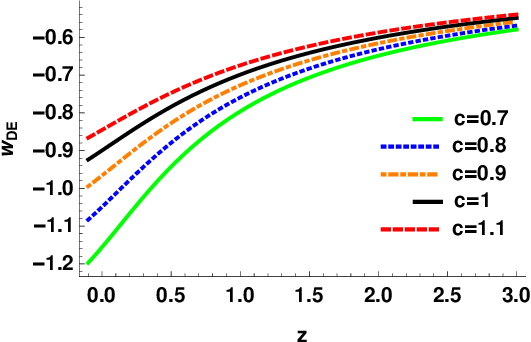}                             
\caption{\it{ The redshift evolution of the equation-of-state parameter $w_{DE}$ of LQG
holographic dark energy, for fixed $q=1.1$ and $\Lambda(\gamma_{0})=1$ and various values of $c$ in units where $M^{2}_{p} =1$. We have imposed $\Omega_{DE}(x = -\ln{(1 + z)} = 0) \equiv \Omega_{DE0}\approx 0.7$ at present in agreement with observations.
}}
\label{hdemultiwdec}
\end{figure} 
 \begin{figure}[!h]
\centering
\includegraphics[width=8cm]{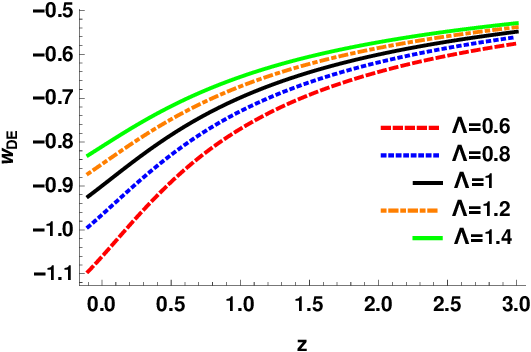}                             
\caption{\it{ The redshift-evolution of the equation-of-state parameter 
$w_{DE}$ of LQG
holographic dark energy, for fixed $q=1.1$ and $c=1$ and various values of $\Lambda(\gamma_0)$ in units where $M^{2}_{p} =1$. We have imposed $\Omega_{DE}(x = -\ln{(1 + z)} = 0) \equiv \Omega_{DE0}\approx 0.7$ at present in agreement with observations.
}}
\label{hdemultiwdeL}
\end{figure} 

Let us now study the effect of the model parameters $q$, $c$ and $\Lambda(\gamma_{0})$ on $w_{DE}$. In Fig. \ref{hdemultiwdeq} we present $w_{DE}$, as a function of the redshift parameter $z$, for various values of the entropic parameter $q$ and for fixed $c=1$ and $\Lambda(\gamma_{0})=1$. As we observe, for increasing values of the entropic parameter $q$, the corresponding evolution of $w_{DE}$, as well as its present value \ie $w_{DE}|_{z = 0}$, tend to lower values. However at times around $z\approx 0.6$ it remains almost unaltered. We mention that for $q\gtrsim 1.9$, the evolution of the equation-of-state parameter at present times ($w_{DE}|_{z = 0}$) always lie in the phantom regime. Thus, according to the value of $q$, the dark energy sector can be quintessence-like, phantom-like, or experience the phantom-divide crossing before or after the present time.
Furthermore, in Fig. \ref{hdemultiwdec} we depict $w_{DE}$ for fixed $q=1.1$ and $\Lambda(\gamma_{0})=1$ and various values of the model parameter $c$. As we can see as $c$  decreases $w_{DE}$, as well as its present value $w_{DE}|(z=0)$, tend to acquire lower values, experiencing the  phantom-divide crossing during the evolution. Additionally note that for $c<0.9$ the value of $w_{DE}|_{z=0}$ lies in the phantom regime. Lastly, Fig. \ref{hdemultiwdeL} depicts the effect of the parameter $\Lambda(\gamma_{0})$ on the evolution of $w_{DE}$, where in this case the parameters $q=1.1$ and $c=1$ are fixed. As we observe in this case, the parameter $\Lambda(\gamma_{0})$ has the same effect as the effect of the parameter $c$ on the evolution of $w_{DE}$, namely as $\Lambda(\gamma_{0})$  decreases, $w_{DE}$, as well as its present value $w_{DE}|(z=0)$, tend to acquire lower values, experiencing the  phantom-divide crossing. In this case we note that for $\Lambda(\gamma_{0})<0.8$, the value of $w_{DE}|_{z=0}$ lies in the phantom regime. In general, the asymptotic value of $w_{DE}$ in the far future, namely at $z\rightarrow -1$, as can be deduced from (\ref{wDE}), it depends on the combination of the parameter values $q$, $c$ and $\Lambda(\gamma_{0})$. In summary, we can see, that  according to the values of the model parameters $q$, $c$ and $\Lambda(\gamma_{0})$, the scenario of LQG holographic dark energy can lead to a very interesting cosmological phenomenology, and can lie in the quintessence or in the phantom regime, or exhibits the phantom-divide crossing during the evolution. 

\section{Conclusions} \label{concl}
In this work, we constructed a new holographic cosmological dark energy scenario, based on Loop Quantum Gravity inspired entropy.  Applying the usual holographic principle at a cosmological framework, imposing the future event 
horizon as the  Infrared  cutoff, but using the Loop Quantum Gravity entropy, instead of the standard  Bekenstein-Hawking one, the former is a non-extensive generalization of Boltzmann-Gibbs entropy, arising from non-extensive statistics and quantum geometry, where the single entropic parameter $q$ quantifies how the probability of frequent events is enhanced relatively to infrequent ones and the deviations from the standard holographic dark energy, possessing it as a particular 
limit, namely for $q\rightarrow 1$. 

Furthermore, we extracted the necessary differential equation that determines the 
evolution of the effective dark energy density, and through which we provided  analytical expressions for 
the corresponding equation-of-state parameter, as well as for the deceleration parameter. As we showed, the scenario at hand can
 successfully describe the thermal history of the Universe, with the the sequence of matter and dark energy eras, while  the transition to acceleration takes place at $z\approx0.6$. Moreover, concerning the dark energy equation-of-state parameter, we showed that according to the values of the three model parameters, namely the Loop Quantum gravity entropic parameter $q$, 
parameter $c$ and the parameter $\Lambda(\gamma_0)$, where $\gamma_0$ is the Barbero-Immirzi parameter, 
the scenario of Loop Quantum gravity holographic dark energy can have a rich cosmological behavior, significally affecting the behavior of the dark energy equation-of-state parameter, which can lie in the quintessence regime, in the phantom regime, or experience the phantom-divide crossing during the evolution.

More specifically, in the case where the model parameters $c$ and $\Lambda(\gamma_0)$ are fixed, increasing $q$ leads to smaller $w_{DE}$ values, while it remains almost unaltered around $z\approx 0.6$. On the other hand, for fixed $q$ and $\Lambda(\gamma_0)$, decreasing $c$ leads $w_{DE}$ to acquire smaller values, from high to low redshifts, while similar behavior we acquire and in the case where we keep the parameters $q$ and $c$ fixed. Finally, in the far future, dark energy dominates completely, and the asymptotic $w_{DE}$ value depends on $q$, $c$ and $\Lambda(\gamma_0)$.

In conclusion, LQG holographic dark energy exhibits more interesting and richer phenomenology in comparison to the standard holographic dark energy, and thus it can be a possible candidate for the description of nature. Definitely, before one considers it as a successful candidate for the description of dark energy, it would be both necessary and interesting to perform a full observational analysis, confronting the scenario with observational data from  Supernovae type Ia (SNIa), Baryonic Acoustic 
Oscillations (BAO), and Cosmic Microwave Background (CMB) probes, Large Scale Structure data. Additionally it is interesting to investigate how the model at hand (and holographic constructions) may alleviate the $H_{0}$ tension too, which as we showed may lead to behavior where $w_{DE}<-1$, which seems to be a requirement in order to $H_{0}$ tension be alleviated \cite{Abdalla:2022yfr}, in order to  extract constraints on the model parameters. These investigations lie beyond the scope of the present work and are left for future projects.

\section{Aknowlegments}
The author is grateful to Emmanuel N. Saridakis for helpful discussion and suggestions on the original manuscript.

\bibliographystyle{JHEP} 
\bibliography{paperf}

\end{document}